\documentclass{article}

\usepackage{PRIMEarxiv}

\usepackage[utf8]{inputenc} 
\usepackage[T1]{fontenc}    
\usepackage{hyperref}       
\usepackage{url}            
\usepackage{booktabs}       
\usepackage{amsfonts}       
\usepackage{nicefrac}       
\usepackage{microtype}      
\usepackage{lipsum}
\usepackage{fancyhdr}       
\usepackage{graphicx}       
\graphicspath{{media/}}     

\title{RUPNet: Residual upsampling network for real-time polyp segmentation
\thanks{\textit{\underline{Citation}}: 
\textbf{Authors. Title. Pages.... DOI:000000/11111.}} 
}

\author{
 Nikhil Kumar Tomar\\
 Machine and Hybrid Intelligence Lab,\\ Department of Radiology,\\ Northwestern University, \\Chicago \\
  USA\\
  \texttt{\{nikhilroxtomar\}@gmail.com} \\
   \And
  Ulas Bagci \\
  Machine and Hybrid Intelligence Lab,\\ Department of Radiology,\\ Northwestern University, \\Chicago \\
  USA\\
  \texttt{\{ulas.bagci\}@northwestern.edu} \\
     \And
Debesh Jha \\
  Machine and Hybrid Intelligence Lab,\\ Department of Radiology,\\ Northwestern University, \\Chicago \\
  USA\\
  \texttt{\{debesh.jha\}@northwestern.edu} \\
}

\begin{document}
\maketitle

\begin{abstract}
Colorectal cancer is among the most prevalent cause of cancer-related mortality worldwide. Detection and removal of polyps at an early stage can help reduce mortality and even help in spreading over adjacent organs. Early polyp detection could save the lives of millions of patients over the world as well as reduce the clinical burden. However, the detection polyp rate varies significantly among endoscopists. There is numerous deep learning-based method proposed, however, most of the studies improve accuracy. Here, we propose a novel architecture,  Residual Upsampling Network (RUPNet) for colon polyp segmentation that can process in real-time and show high recall and precision. The proposed architecture, RUPNet, is an encoder-decoder network that consists of three encoders, three decoder blocks, and some additional upsampling blocks at the end of the network. With an image size of $512 \times 512$, the proposed method achieves an excellent real-time operation speed of 152.60 frames per second with an average dice coefficient of 0.7658,  mean intersection of union of 0.6553, sensitivity of 0.8049, precision of 0.7995, and accuracy of 0.9361. The results suggest that RUPNet can give real-time feedback while retaining high accuracy indicating a good benchmark for early polyp detection.
\end{abstract}


\section{Introduction}
Colonoscopy is considered the gold standard and is recommended screening tool for colon cancer diagnosis and follow-up. Early detection and removal of adenomas is crucial in colonoscopy for removing incidence and mortality. A 1\% increase in adenoma detection rate showed associated with a 3\% decrease in interval colorectal cancer incidence~\cite{urban2018deep}. Still, the adenoma miss rate is around 6-27\%~\cite{ahn2012miss}. Therefore, an endoscopist is expected to perform a high-quality colonoscopy examination. The essential quality indicators for colonoscopy are adenoma detection rate (ADR) and withdrawal time (WT)~\cite{rex2002quality,kaminski2010quality}. Once the adenomas are detected during examination there are chances that the endoscopists become less motivated in finding the subsequent adenomas~\cite{lai2009boston,imperiale2016new,amano2018number}.

There are various reasons for high polyp miss-rate such as size, shape, quality of bowel preparation, and faster colonoscope withdrawal time~\cite{ahn2012miss}. Sometimes during a routine examination, the polyp is not recognized and sometimes despite being recognized they are missed. The polyp detection rate can be improved by detecting the unrecognized polyps that are on the visual field~\cite{mahmud2015computer}. Therefore, an algorithm-based second observer might help to improve the polyp detection rate during live colonoscopy examination. Additionally, a real-time computer-aided diagnosis system could highlight the suspicious polyps that could be clinically relevant for selecting optimal treatment, ideal colonoscopic resection and reducing overall cost~\cite{rex2017colorectal}. 

Deep learning based algorithms can highlight the presence of precancerous tissue in the colon and have the potential to improve the diagnostic performance of endoscopists.  In clinical practice, precise polyp segmentation provides important information in the early detection of colorectal cancer. Despite of success of deep learning algorithms for automatic polyp segmentation, it is still an unsolved problem. This is because even the most successful method fails when tested across new datasets obtained from the other centers mostly because of different imaging protocols, different cohorts populations and various scanners. The challenges which make polyp segmentation complicated are the variable nature of polyps in their size, shapes, subtle appearances (for example, sessile serrated lesions) and high recurrence rate etc. The human factors such as bowel preparation and skill of the particular endoscopist can also affect examination procedure~\cite{jha2022machine}.

Tomar et al.~\cite{tomar2022fanet} proposed a feedback attention network (FANet) for improved biomedical image segmentation, where they showed the state-of-the-art performance on seven publicly available benchmark datasets. FANet unifies the mask of the previous epoch with the current training epoch and rectifies the prediction iteratively during test time for improved performance. Recently, Biffi et al.~\cite{biffi2022novel} proposed an intelligent medical device for real-time optical characterization of colonoscopy polyps which can be also integrated easily into clinics. Besides that, there are several recent works on colonoscopy targeting polyps attributes such as size and count~\cite{tomar2022tganet}, flat, sessile or diminutive polyps~\cite{jha2021comprehensive}, out-of-distribution polyp detection through cross dataset test~\cite{srivastava2022gmsrf,jha2021comprehensive}, and synthetic data generation for improving the performance of the network~\cite{fagereng2022polypconnect}. As the study of colonoscopy cancer is becoming mature more emphasis is being given towards exploring the possibilities of new technological advancement such as diagnostic classification, risk stratification, and outcome examination~\cite{diao2022computer}. 

The main contribution are as follow:
\begin{enumerate}
\item We propose RUPNet, a novel lightweight encoder decoder based architecture for real-time polyp segmentation. Our algorithm is based on residual block and the architecture is designed in a way that requires low computational power for training.  

\item We evaluate RUPNet with the popular benchmarks and showed improvement over those architectures. Most interestingly, our method obtained a real-time speed of 152.60 which is important for the clinical integration. 
\end{enumerate}

\vspace{5 mm}

The rest of the paper is organized as follows. In Section~\ref{sec:method}, we present the method. In Section~\ref{sec:experimental_setup}, we present the experimental setup. Then, in Section~\ref{sec:results}, we present results and Discussion. Finally, we conclude our work in Section~\ref{sec:conclusion}.

\vspace{3mm}
\section{Method}
\begin{figure}[!t]
    \centering
    \includegraphics[width = 0.6\textwidth]{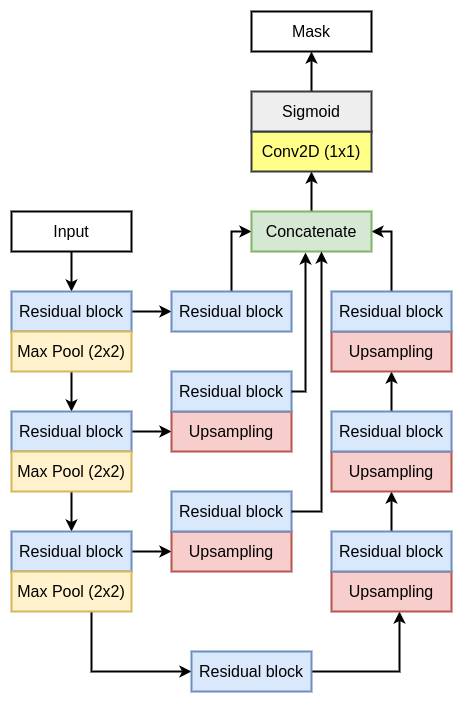}
    \caption{Overall architecture of the RUPNet.}
    \label{fig:rupnet}
\end{figure}

\vspace{3mm}
\label{sec:method}
Figure~\ref{fig:rupnet} shows the block diagram of the proposed RUPNet architecture. RUPNet is an encoder-decoder network that consists of three encoder blocks, three decoder blocks, and a few additional upsampling blocks at the end of the network. The input image is passed to the first encoder block consisting of the residual block, followed by $2\times2$ max-pooling. The residual block~\cite{he2016deep} helps propagate information over layers that allows to build a deeper neural network for solving the degradation problems in each encoder block. It helps to improve the channel inter-dependencies and reduces the overall computational cost. The max-pooling is an operation for computing the maximum value for patches of feature maps which is used for creating downsampled feature maps. The output of the last encoder block passes through a residual block, which acts as the bridge between the encoder network and the decoder network. The output of the bridge acts as the input for the decoder network.

The decoder begins with a bilinear upsampling with a factor of two. After that, it is followed by a residual block. The feature map increases in height and width with every decoder block until it reaches the last decoder. Here, we take the skip connections, followed by a bilinear upsampling to ensure that all the skip connections have the same resolution. The skip connection in the network allows the passing of lower-level semantic information to the latter layer, which helps preserve the information for better information flow. Next, we concatenate the upsampled feature maps with the output of the last decoder block. Finally, we have a $1\times1$ convolution with the Sigmoid activation function to get the final mask.

 \begin{figure}[!t]
    \centering
    \includegraphics[width = 0.9\textwidth]{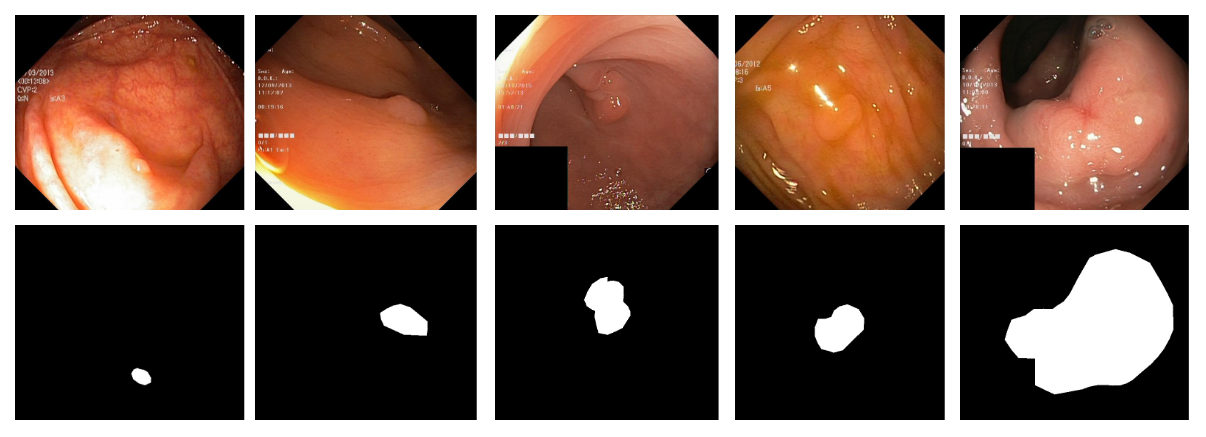}
    \caption{The figure shows examples of Kvasir-SEG~\cite{jha2020kvasir} used for training the RUPNet algorithm.}
    \label{fig:kvasir}
\end{figure}

\section{Experimental setup}
\label{sec:experimental_setup}

We have selected the Kvasir-SEG~\cite{jha2020kvasir} dataset consisting of 1,000 polyp images and their corresponding ground truth for training the proposed model. Figure~\ref{fig:kvasir} shows the example images from the Kvasir-SEG dataset. We use 880 images and masks for training our model and the rest 120 images for testing. We have performed extensive data augmentation, such as flipping, rotation, random brightness, etc., to increase the number of training samples. Our experiments are run on an NVIDIA RTX 3090 GPU system. We have used the Adam optimizer having a learning rate of 1e$^{-4}$  and a batch size of 8. We have used a combination of binary cross-entropy and dice loss as a loss function. For the evaluation, we use standard medical image segmentation based metrics such as mean intersection over union (mIoU), dice coefficient (mDSC), recall or sensitivity, precision, accuracy, F2-score and processing speed. 
\begin{table*}[t!]
   \def\arraystretch{1.6}
    \setlength\tabcolsep{6pt}
    \par\bigskip
\centering
\caption{Quantitative results on the Kvasir-SEG~\cite{jha2020kvasir} test dataset.}
 \begin{tabular} {l|c|c|c|c|c|c}
\toprule
\textbf{Method} & \textbf{mDSC}  &\textbf{mIoU}  &\textbf{Recall}& \textbf{Precision} &\textbf{Accuracy} &\textbf{FPS}\\ 
\hline

U-Net~\cite{ronneberger2015u}  &0.5969  & 0.4713 &0.6171 & 0.6722 &0.8936 & 22.02 \\

ResUNet~\cite{zhang2018road}&0.6902 &0.5721 &0.7248 &0.7454 &0.9169 &14.82\\

ResU-Net++~\cite{jha2019resunet++}&0.7143 &0.6126 &0.7419 &0.7836 &0.9172 &7.01\\

RUPNet (Ours) &\textbf{0.7658} &\textbf{0.6553} &\textbf{0.8049} &\textbf{0.7995} &\textbf{0.9361} &\textbf{152.60}\\

\bottomrule
\end{tabular}
\label{tab:results}
\end{table*}

\begin{figure}[!t]
\centering
\includegraphics[width=0.6\linewidth]{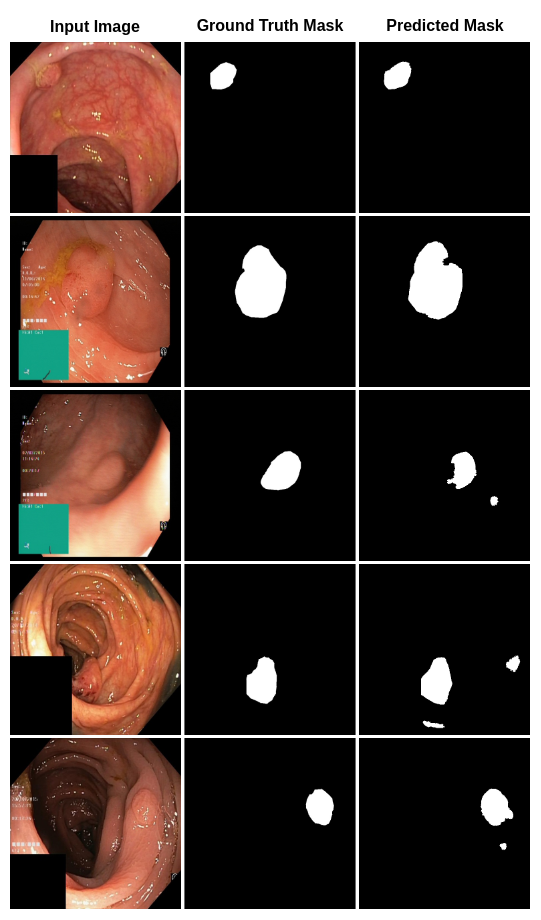} 
\caption{The figure shows the example of qualitative results of RUPNet prediction. From the example, it can be observed that RUPNet can accurately segment medium and smaller-sized polyps. However, it also shows over-segmentation when exposed to light lighting conditions.}
\label{fig:qualitative}
\end{figure}

\section{Result and Discussion}
\label{sec:results}
\subsection{Comparison with State-of-the-Art: } 
Table~\ref{tab:results} shows the result of the RUPNet. The proposed architecture obtain an F1-score of 0.7658, mIoU of 0.6553,  recall of 0.8049, precision of 0.7995, and F2-score 0.9361. With the image resolution of $512\times512$, the proposed method achieves a real-time operation speed of 152.60 frames per second (FPS). The main reason behind higher speed is its architectural design, which has only a few trainable parameters, making the network lightweight for real-time processing. The most competitive benchmarking method of RUPNet was ResUNet++ with a dice coefficient of 0.7143. However, it has only a speed of 7.0193 FPS. Therefore, our method is 21.74 times faster than one of the widely used benchmark in the field of automated polyp segmentation.

Figure~\ref{fig:qualitative} shows the qualitative results of RUPNet. The qualitative result shows that RUPNet can correctly segment smaller and medium-sized polyps that are commonly missed due to their size. The visual results also show that RUPNet shows over-segmentation for the lighting conditions, as evidenced last two examples from Figure~\ref{fig:qualitative}. Thus, both the qualitative and quantitative results exhibit an acceptable overall performance. 

\subsection{Limitations and open challenges}
Our study has some limitations. We classify images into polyp and non-polyp regions. However, we do not cover if the polyp is adenoma or non-adenoma and does not look into diagnostic classes such as hyperplastic or sessile polyps. For the algorithm to be integrated into the clinical workflow,  the proposed algorithm should perform well on out-of-the-distribution datasets that are collected from different medical centres. Additionally, the algorithm should perform well in conditions such as camouflage and noise and should possess real-time processing speed. Although, we achieve real-time speed our algorithm is trained and tested on the same distribution datasets. We have not experimented on multi-center datasets, and our algorithm shows that there are still some challenges when there is camouflage which is caused by the false positive. This is also observed through the qualitative analysis. Moreover, we have not performed statistical tests for any experiments, which is also a way of interpreting the best model.

\section{Conclusion}
In this work, we proposed the RUPNet architecture that utilizes a simple residual block to accurately segment polyps with high processing speed. The experimental results on the colon polyp dataset showed that RUPNet can accurately give real-time feedback with high accuracy, and it might help endoscopists in early polyp detection in clinics. In the future, we plan to extend RUPNet by adding transformer blocks in the encoder and evaluate with the multi-centre out-of-distribution dataset to study the robustness of our algorithm. Additionally, we will also research on distinguishing colon polyps based on their attributes, such as the number of polyps counts (for example, one or many) and their size (for example, $\leq 5 mm$, between $5 mm \leq 10 mm$, and $\geq$ than 10 mm) and compare it with the recent state-of-the-art polyp segmentation benchmark. 

\section*{Acknowledgments}
This project is supported by the NIH funding: R01-CA246704 and R01-CA240639.

\bibliographystyle{unsrt}  
\bibliography{references}

\begin{thebibliography}{10}

\bibitem{urban2018deep}
Gregor Urban, Priyam Tripathi, Talal Alkayali, Mohit Mittal, Farid Jalali,
  William Karnes, and Pierre Baldi.
\newblock Deep learning localizes and identifies polyps in real time with 96\%
  accuracy in screening colonoscopy.
\newblock {\em Gastroenterology}, 155(4):1069--1078, 2018.

\bibitem{ahn2012miss}
Sang~Bong Ahn, Dong~Soo Han, Joong~Ho Bae, Tae~Jun Byun, Jong~Pyo Kim, and
  Chang~Soo Eun.
\newblock The miss rate for colorectal adenoma determined by quality-adjusted,
  back-to-back colonoscopies.
\newblock {\em Gut and liver}, 6(1):64, 2012.

\bibitem{rex2002quality}
Douglas~K Rex, John~H Bond, Sidney Winawer, Theodore~R Levin, Randall~W Burt,
  David~A Johnson, Lynne~M Kirk, Scott Litlin, David~A Lieberman, Jerome~D
  Waye, et~al.
\newblock Quality in the technical performance of colonoscopy and the
  continuous quality improvement process for colonoscopy: recommendations of
  the us multi-society task force on colorectal cancer.
\newblock {\em Official journal of the American College of Gastroenterology|
  ACG}, 97(6):1296--1308, 2002.

\bibitem{kaminski2010quality}
Michal~F Kaminski, Jaroslaw Regula, Ewa Kraszewska, Marcin Polkowski, Urszula
  Wojciechowska, Joanna Didkowska, Maria Zwierko, Maciej Rupinski, Marek~P
  Nowacki, and Eugeniusz Butruk.
\newblock Quality indicators for colonoscopy and the risk of interval cancer.
\newblock {\em New England journal of medicine}, 362(19):1795--1803, 2010.

\bibitem{lai2009boston}
Edwin~J Lai, Audrey~H Calderwood, Gheorghe Doros, Oren~K Fix, and Brian~C
  Jacobson.
\newblock The boston bowel preparation scale: a valid and reliable instrument
  for colonoscopy-oriented research.
\newblock {\em Gastrointestinal endoscopy}, 69(3):620--625, 2009.

\bibitem{imperiale2016new}
Thomas~F Imperiale and Douglas~K Rex.
\newblock A new quality indicator of colonoscopy: caveat emptor.
\newblock {\em Gastrointestinal Endoscopy}, 84(3):507--511, 2016.

\bibitem{amano2018number}
Takahiro Amano, Tsutomu Nishida, Hiromi Shimakoshi, Akiyoshi Shimoda, Naoto
  Osugi, Aya Sugimoto, Kei Takahashi, Kaori Mukai, Dai Nakamatsu, Tokuhiro
  Matsubara, et~al.
\newblock Number of polyps detected is a useful indicator of quality of
  clinical colonoscopy.
\newblock {\em Endoscopy International Open}, 6(07):E878--E884, 2018.

\bibitem{mahmud2015computer}
Nadim Mahmud, Jonah Cohen, Kleovoulos Tsourides, and Tyler~M Berzin.
\newblock Computer vision and augmented reality in gastrointestinal endoscopy.
\newblock {\em Gastroenterology report}, 3(3):179--184, 2015.

\bibitem{rex2017colorectal}
Douglas~K Rex, C~Richard Boland, Jason~A Dominitz, Francis~M Giardiello,
  David~A Johnson, Tonya Kaltenbach, Theodore~R Levin, David Lieberman, and
  Douglas~J Robertson.
\newblock Colorectal cancer screening: recommendations for physicians and
  patients from the us multi-society task force on colorectal cancer.
\newblock {\em Gastroenterology}, 153(1):307--323, 2017.

\bibitem{jha2022machine}
Debesh Jha.
\newblock Machine learning-based classification, detection, and segmentation of
  medical images.
\newblock {\em PhD thesis}, 2022.

\bibitem{tomar2022fanet}
Nikhil~Kumar Tomar, Debesh Jha, Michael~A Riegler, H{\aa}vard~D Johansen, Dag
  Johansen, Jens Rittscher, P{\aa}l Halvorsen, and Sharib Ali.
\newblock Fanet: A feedback attention network for improved biomedical image
  segmentation.
\newblock {\em IEEE Transactions on Neural Networks and Learning Systems},
  2022.

\bibitem{biffi2022novel}
Carlo Biffi, Pietro Salvagnini, Nhan~Ngo Dinh, Cesare Hassan, Prateek Sharma,
  and Andrea Cherubini.
\newblock A novel ai device for real-time optical characterization of
  colorectal polyps.
\newblock {\em NPJ digital medicine}, 5(1):1--8, 2022.

\bibitem{tomar2022tganet}
Nikhil~Kumar Tomar, Debesh Jha, Ulas Bagci, and Sharib Ali.
\newblock Tganet: Text-guided attention for improved polyp segmentation.
\newblock In {\em Proceedings of the International Conference on Medical image
  computing and computer-assisted intervention (MICCAI)}, 2022.

\bibitem{jha2021comprehensive}
Debesh Jha, Pia~H Smedsrud, Dag Johansen, Thomas de~Lange, H{\aa}vard~D
  Johansen, P{\aa}l Halvorsen, and Michael~A Riegler.
\newblock {A comprehensive study on Colorectal Polyp Segmentation With}
  {ResUNet++}, {Conditional Random Field and Test-Time Augmentation}.
\newblock {\em IEEE Journal of Biomedical and Health Informatics},
  25(6):2029--2040, 2021.

\bibitem{srivastava2022gmsrf}
Abhishek Srivastava, Sukalpa Chanda, Debesh Jha, Umapada Pal, and Sharib Ali.
\newblock {GMSRF-Net}: {An improved generalizability with global multi-scale
  residual fusion network for polyp segmentation}.
\newblock In {\em Proceedings of the International Conference on Pattern
  Recognition (ICPR)}, pages 4321--4327, 2022.

\bibitem{fagereng2022polypconnect}
Jan~Andre Fagereng, Vajira Thambawita, Andrea~M Stor{\aa}s, Sravanthi Parasa,
  Thomas de~Lange, P{\aa}l Halvorsen, and Michael~A Riegler.
\newblock Polypconnect: Image inpainting for generating realistic
  gastrointestinal tract images with polyps.
\newblock {\em arXiv preprint arXiv:2205.15413}, 2022.

\bibitem{diao2022computer}
James~A Diao and Joseph~C Kvedar.
\newblock Computer copilots for endoscopic diagnosis, 2022.

\bibitem{he2016deep}
Kaiming He, Xiangyu Zhang, Shaoqing Ren, and Jian Sun.
\newblock Deep residual learning for image recognition.
\newblock In {\em Proceedings of the IEEE conference on computer vision and
  pattern recognition (CVPR)}, pages 770--778, 2016.

\bibitem{jha2020kvasir}
Debesh Jha, Pia~H Smedsrud, Michael~A Riegler, P{\aa}l Halvorsen, Thomas~de
  Lange, Dag Johansen, and H{\aa}vard~D Johansen.
\newblock {Kvasir-SEG:} a segmented polyp dataset.
\newblock In {\em Proceedings of the International Conference on Multimedia
  Modeling (MMM)}, pages 451--462, 2020.

\bibitem{ronneberger2015u}
Olaf Ronneberger, Philipp Fischer, and Thomas Brox.
\newblock {U-Net:} {Convolutional Networks for Biomedical Image Segmentation}.
\newblock In {\em Proceedings of the International Conference on Medical image
  computing and computer-assisted intervention (MICCAI)}, pages 234--241, 2015.

\bibitem{zhang2018road}
Zhengxin Zhang, Qingjie Liu, and Yunhong Wang.
\newblock Road extraction by deep residual u-net.
\newblock {\em IEEE Geoscience and Remote Sensing Letters}, 15(5):749--753,
  2018.

\bibitem{jha2019resunet++}
Debesh Jha, Pia~H Smedsrud, Michael~A Riegler, Dag Johansen, Thomas De~Lange,
  P{\aa}l Halvorsen, and H{\aa}vard~D Johansen.
\newblock Resunet++: An advanced architecture for medical image segmentation.
\newblock In {\em Proceedings of the International Symposium on Multimedia
  (ISM)}, pages 225--2255, 2019.

\end{thebibliography}

\end{document}